\documentclass[12pt]{article}
\usepackage[utf8]{inputenc}
\usepackage{threeparttable}
\title{Note}

\usepackage{natbib, url}
\usepackage{graphicx}
\usepackage{lscape, upgreek, prodint}
\usepackage[colorlinks=true, citecolor=blue, linkcolor=blue, urlcolor=blue]{hyperref}
\usepackage{verbatim, color,amsmath,amssymb, epsfig, listings}
\def\log{\hbox{log}}

\def\n{\nonumber}
\def\boxit#1{\vbox{\hrule\hbox{\vrule\kern6pt
 \vbox{\kern6pt#1\kern6pt}\kern6pt\vrule}\hrule}}

\def\bse{\begin{eqnarray*}}
\def\ese{\end{eqnarray*}}
\def\be{\begin{eqnarray}}
\def\ee{\end{eqnarray}}
\def\bq{\begin{equation}}
\def\eq{\end{equation}}
\def\bse{\begin{eqnarray*}}
\def\ese{\end{eqnarray*}}
\def\pr{\hbox{pr}}

\def\wh{\widehat}
\newtheorem{Th}{\underline{\bf Theorem}}

\newtheorem{proposition}{Proposition}

\newtheorem{Exa}{Example}

\def\boxit#1{\vbox{\hrule\hbox{\vrule\kern6pt
         \vbox{\kern6pt#1\kern6pt}\kern6pt\vrule}\hrule}}

\def\wh{\widehat}

\def\log{\hbox{log}}

\def\bse{\begin{eqnarray*}}
\def\ese{\end{eqnarray*}}
\def\be{\begin{eqnarray}}
\def\ee{\end{eqnarray}}
\def\bq{\begin{equation}}
\def\eq{\end{equation}}
\def\bse{\begin{eqnarray*}}
\def\ese{\end{eqnarray*}}
\def\pr{\hbox{pr}}
\def\wh{\widehat}

\def\log{{\rm log}}

\def\0{{\bf 0}}

\def\t{{\bf t}}

\usepackage{titlesec}
\makeatother 
\usepackage{tcolorbox}

\definecolor{codegreen}{rgb}{0,0.6,0}
\definecolor{codegray}{rgb}{0.5,0.5,0.5}
\definecolor{codepurple}{rgb}{0.58,0,0.82}
\definecolor{backcolour}{rgb}{0.95,0.95,0.92}

\lstdefinestyle{mystyle}{
    commentstyle=\color{magenta},
    keywordstyle=\color{blue},
    stringstyle=\color{purple},
    basicstyle=\ttfamily\small,
    breakatwhitespace=false,         
    breaklines=true,                 
    captionpos=b,                    
    tabsize=2
}

\begin{document}

\begin{center}
{\Large{\bf  }}
\end{center}
\vskip 5mm \baselineskip=18pt
\begin{center}
{\bf \Large   Efficient Estimation of the Additive Risks Model for Interval-Censored Data}
\end{center}

\thispagestyle{empty}
\vskip 5mm
\begin{center}
{\sc  Tong Wang$^{1, 3}$,  Dipankar Bandyopadhyay$^2$ and  Samiran Sinha$^{3, \dagger}$} \\
\vskip 3mm 
$^1$
School of Statistics and Data Science, Nankai University, Tianjin, China\\
$^2$ Department of Biostatistics, Virginia Commonwealth University, Richmond, VA, USA\\
 $^{3}$ Department of Statistics, Texas A\&M University, College Station, TX, USA \\
$^\dagger$email:  sinha@stat.tamu.edu\\
\end{center}

\vskip 5mm

\begin{center}
Abstract
\end{center}
In contrast to the popular Cox model which presents a multiplicative covariate effect specification on the time to event hazards, the semiparametric additive risks model (ARM) offers an attractive additive specification, allowing for direct assessment of the changes or the differences in the hazard function for changing value of the covariates. The ARM is a flexible model, allowing the estimation of both time-independent and time-varying covariates. It has a nonparametric component and a regression component identified by a finite-dimensional parameter. This chapter presents an efficient approach for maximum-likelihood (ML) estimation of the nonparametric and the finite-dimensional components of the model via the minorize-maximize (MM) algorithm for case-II interval-censored data. The operating characteristics of our proposed MM approach are assessed via simulation studies, with illustration on a breast cancer dataset via the \texttt{R} package \texttt{MMIntAdd}. It is expected that the proposed  computational approach will not only provide scalability to the ML estimation scenario but may also simplify the computational burden of other complex likelihoods or models.

\vskip 8mm 


\baselineskip=24pt
\allowdisplaybreaks

\newpage

\setcounter{page}{1}

\allowdisplaybreaks

\baselineskip=24pt

\section{Introduction}

Interval-censoring \citep{Boga2018}, which occurs when the failure time is only known to lie in an interval instead of being observed precisely, abounds in demographical, sociological, and biomedical studies \citep{zhang2010interval}. There are broadly two main types of interval-censored data: case-I and case-II interval-censored data. Case-I interval-censored data, also called current status data \citep{Martinussen2002}, is not the focus of this chapter. Here, we focus on case-II interval censoring, where the time to events are a mixture of left-, right-, and interval-censoring. 
Specifically, case-2 interval-censored data consists of some left-censored time-to-events, some right-censored time to-events, and some interval-censored time-to-events, and the proportion of interval-censored time-to-events never goes to zero as the sample size increases. 
This work aims to present an efficient algorithm for maximum likelihood (ML) estimation of the additive risks model \citep{lin1994semiparametric}, henceforth ARM, for the case-II interval-censored data.

The ARM is specified by the hazard function
\be\label{eqm1}
h(t|X(t))=\lambda(t)+\beta^\top X(t),
\ee
where, $X(t)$ denotes a vector of possibly  time-dependent covariate, $\beta$ is the corresponding regression parameter, 
and $\lambda(t)$  is the baseline hazard function. In this model, the effect of a covariate  can be measured via the difference in the hazard function 
 for different covariate values at any given time. 
In (\ref{eqm1}), the effect of a covariate is assumed to be constant on the hazard function. However, it can be relaxed to any known parametric form that is possibly time-dependent. \cite{Lin1994} used this ARM to analyze right-censored data. Under case-II interval-censoring,
\cite{Zeng2006} proposed an ML method to estimate both the baseline hazard function and regression parameters of the model. In contrast,  \cite{Wang2010} considered a martingale-based estimation procedure, focusing only on the estimation of the regression parameters bypassing baseline hazard estimation -- a critical component to study the event of interest. Furthermore,
\cite{Martinussen2002} and \cite{Wang2020Korean} proposed to use a sieve ML approach to model the baseline hazard $\lambda(t)$ under current status and case-II interval-censoring, respectively. The sieve method requires an appropriate choice of the sieve parameter space and the number of knots.

In our ML approach of fitting the ARM  to the interval-censored data, the baseline survival function was modeled as a nonparametric step function with a jump at the observed inspection time points. The computation of the ML estimates through direct maximization of the observed data likelihood function is problematic due to a large number of parameters. Note, although the regression parameter is finite-dimensional, the baseline hazard function contributes a large number of parameters that tend to increase with the sample size when the inspection time is continuous \citep{Zeng2006}. To circumvent this computational difficulty in high-dimensional ML maximization, we develop a novel Minorize-Maximization (MM) algorithm \citep{Hunter2004,Wu2010}. The proposed method can handle both time-independent and time-dependent covariates.  By applying this technique, the original high-dimensional optimization problem reduces to a simple Newton-Raphson update of the parameters. Moreover, in each step of the Newton-Raphson method, we do not need to invert any high-dimensional matrix. All these are possible with a clever choice of the surrogate function, and details of this choice are discussed in the next section. Extensive simulation studies confirm that the proposed MM algorithm can estimate the parameters adequately, with a significantly reduced computation time than direct maximization.

The efficiency of an MM algorithm relies on choosing an appropriate minorizing function that requires understanding and applying mathematical inequalities in the right places. MM algorithms have been developed  
in quantile regression \citep{Hunter2000}, variable selection \citep{Hunter2005}, and in various areas of 
machine-learning; see the review article by \cite{Nguyen2017}, and the references therein. This algorithm has been used in analyzing censored time-to-event data with the proportional odds model \citep{Hunter2002}, clustered time-to-event data with the Gamma frailty model \citep{Huang2019}, and recently in analyzing clustered current status data with the generalized odds ratio model \cite{TongWang2020}. This book chapter presents our maiden attempt to employ the MM algorithm for inference under the ARM for interval-censored data to the best of our knowledge.  
 The novelty of the work lies in developing an efficient ML estimation procedure for this semiparametric ARM for analyzing case-II interval-censored data. For the consistency and asymptotic normality of the ML estimator, we refer to \cite{Zeng2006}.

The remainder of the chapter is organized as follows. 
After specifying the notations and hazard specifications, Section \ref{sec:model} presents the likelihood of our proposed ARM. Section \ref{sec:MM} presents the relevant details of the proposed MM algorithm, including variance estimation, and complexity analysis.  The finite-sample performances of our estimators are evaluated via simulation studies using synthetic data in Section \ref{sec:simulation}. Section \ref{sec:realdata} illustrates our proposed methodology via application to a well-known breast cosmesis data with interval-censored endpoints. Relevant model-fitting and implementation using our \texttt{R} package \texttt{MMIntAdd} are presented in Section \ref{sec:forR}. Finally, Section \ref{sec:conclusion} concludes, alluding to some future work.

\section{Statistical Model} \label{sec:model}

\subsection{Notations and Setup}

Let $T_i$ denote the  time-to-event for the $i$th subject. Our observed interval-censored data from
$n$ independent subjects are given by $\{L_i,R_i,X_i,\Delta_{L,i},\Delta_{I,i},\Delta_{R,i}\}$, $i=1, \dots, n$, where $L_{i}$ and $R_{i}$ are left- and right-endpoints of the intervals, $X_i$ is a $p\times 1$ vector of time-dependent covariates, and $\Delta_{L,i}$, $\Delta_{I,i}$ and $\Delta_{R,i}$ represent the left-, interval-, and right-censoring indicators, respectively. If  $T_i$ is left-censored, then  $T_i$ falls in $(0, L_i]$ and $\Delta_{L, i}=1$ while $\Delta_{I, i}=\Delta_{R, i}=0$.  If  $T_i$  is interval-censored, then $T_i$ falls in $(L_i, R_i]$ and $\Delta_{L, i}=\Delta_{R, i}=0$ while $\Delta_{I, i}=1$. Finally, if $T_i$ is right censored, then  $T_i$ falls in $(R_i, \infty)$ and $\Delta_{L, i}=\Delta_{I, i}=0$ while $\Delta_{R, i}=1$. As a placeholder, we can set 
$R_i$ to any number larger than $L_i$ for left censored time-to-event, and $L_i$ to any number smaller than $R_i$ for right-censored time-to-event.

With the hazard function of the ARM given in (\ref{eqm1}), 
 the cumulative hazard is 
$H(t; X)= \Lambda(t)+ \beta^\top Z_x(t)$, where $\Lambda(t)=\int_0^t \lambda(s)ds$ and $Z_x(t)=\int^t_0 X(s)ds$. When the covariate is time independent,   
$Z_x(t)=\int^t_0 X(s) ds= X t$. Given the covariates, the survival probability  is 
\bse
S(t; X)=\exp[-\{\Lambda(t)+\beta^\top Z_x(t)\}]. 
\ese
For the nonparametric ML estimation, assume that $\Lambda(t)$ is a step function with jump $\lambda_k$ at $t_k\;(k=0,\ldots,m)$, i.e., $\Lambda(t)=\sum_{k: t_k\le t}\lambda_k$, where $t_1<\cdots<t_m$,  denote the unique inspection time points. In the example below, we further illustrate the calculation of $\Lambda(t)$ for the interval-censored scenario.

\begin{Exa}
Consider a hypothetical dataset with interval-censored time to events from eight subjects,  $(0, 0.5]$, $(0,5]$, $(2,5]$, $(1,2.5]$, $(1.5,2.25]$, $(3,4.2]$, $(2,\infty)$, $(3.2,\infty)$, where the first two  are left-censored, the next four  are interval-censored and the last two  are right-censored. Then the unique inspection time points $(t_1,t_2,\ldots,t_{10})^\top =(0.5,1,1.5,2,2.25,2.5,3,3.2,4.2,5)^\top $. 
  Let $(\lambda_1,\lambda_2,\ldots,\lambda_{10})^\top $ are the jumps corresponding to $t$'s.  Then $\Lambda(1.75)=\lambda_1+\lambda_2+\lambda_3$ and likewise $\Lambda(3.5)=\lambda_1+\cdots+\lambda_7+\lambda_8$.
\end{Exa}

\subsection{Likelihood} It is assumed that that distribution of the window of the inspection time $(L, R)$ is independent of the time-to-event $T$, and the support of $(L, R)$ is $\Omega=\{(l, r): 0< l_0\leq l<r\leq r_0<\infty\}$. 
The density function  of $(L, R)$ is assumed to be positive over $\Omega$  
and $\pr(T<l_0|X)$ and $\pr(T>r_0|X)$ have a positive lower bound that is  strictly greater than zero.
Like \cite{Zeng2006}, $\beta$ is assumed to lie in a compact set of multidimensional Euclidean space, $\Lambda(0)=0$ and $\Lambda(t)>0$ is assumed to be a non-decreasing function, and  the covariates are assumed to lie in a compact set of multidimensional Euclidean space. 
 Let $\lambda=(\lambda_1,\ldots,
\lambda_m)^\top $, then the  observed likelihood and the log-likelihood functions are  
\bse
\mathcal{L}(\lambda,\beta)=\prod_{i=1}^n\{1-S(L_i; X_i)\}^{\Delta_{L,i}}\{S(L_i; X_i)-S(R_i; X_i)\}^{\Delta_{I,i}}\{S(R_i; X_i)\}^{\Delta_{R,i}}, 
\ese
and 
\begin{eqnarray}
\ell(\lambda,\beta)&=&\sum_{i=1}^n\biggl[\Delta_{L,i}\log\{1-S(L_i; X_i)\}+\Delta_{I,i}\log\{S(L_i; X_i)-S(R_i; X_i)\}+\Delta_{R,i}\log\{S(R_i; X_i)\}\biggl]\nonumber\\
&=&\sum_{i=1}^n\biggl[
\Delta_{L,i}\log\{1-S(L_i; X_i)\}+\Delta_{I,i}\log\{S(L_i; X_i)\}+\Delta_{I,i}\log\{1-S^{-1}(L_i; X_i)S(R_i; X_i)\}\nonumber\\
&& +\Delta_{R,i}\log\{S(R_i,X_i)\}
\biggl]\nonumber\\
&=& \ell_1(\lambda,\beta)+\ell_2(\lambda,\beta)+\ell_3(\lambda,\beta)+\ell_4(\lambda,\beta),\label{log-likelihood}
\end{eqnarray}
where 
\bse
\ell_1(\lambda,\beta)&=&\sum_{i=1}^n\Delta_{L,i}\log\{1-S(L_i|X_i)\}
=\sum_{i=1}^n\Delta_{L,i}\log[1-\exp\{-
\sum_{k: t_k\le L_i}\lambda_k-\beta^\top Z_{x_i}(L_i)\}],\\
\ell_2(\lambda,\beta)&=&\sum_{i=1}^n\Delta_{I,i}\log\{S(L_i|X_i)\}=-\sum_{i=1}^n\Delta_{I,i}\left\{\sum_{k: t_k\leq L_i}\lambda_k+\beta^\top Z_{x_i}(L_i)\right\},\\
\ell_3(\lambda,\beta)&=&\sum_{i=1}^n\Delta_{I,i}\log\{1-S^{-1}(L_i|X_i)S(R_i|X_i)\}\\
&=&\sum_{i=1}^n\Delta_{I,i}\log\Bigg(1-\exp\bigg[-\sum_{k: L_i<t_k\leq R_i}\lambda_k-\beta^\top \{Z_{x_i}(R_i)-Z_{x_i}(L_i)\}\bigg]\Bigg),\\
\ell_4(\lambda,\beta)&=&\sum_{i=1}^n\Delta_{R,i}\log\{S(R_i|X_i)\}=-\sum_{i=1}^n\Delta_{R,i}\left\{\sum_{k:t_k\leq R_i}\lambda_k+\beta^\top Z_{x_i}(R_i)\right\}.
\ese
It is understood that maximization of $\ell(\lambda, \beta)$ is not straight-forward due to the presence of $\lambda$ and $\beta$ in a non-separable functional form. Therefore, 
in the next section, we develop an efficient optimization technique aided by the MM algorithm to estimate $\lambda$ and $\beta$.

\section{Estimation} 
\subsection{MM algorithm}\label{sec:MM}
For developing a computationally efficient  MM algorithm, 
we need to find a suitable minorization function. To develop 
such a minorization function,  we use  a result from the recent literature \citep{TongWang2020} along with some standard mathematical inequalities. Define $\lambda_0=(\lambda_{10},\ldots,\lambda_{m0})^\top $ and   $u_0(L_i,X_i)=\sum_{k: t_k\leq L_i}\lambda_{k0}+\beta_0^\top Z_{x_i}(L_i)$, 
 $u_0(R_i,X_i)=\sum_{k: t_k\leq R_i}\lambda_{k0}+\beta^\top_0Z_{x_i}(R_i)$ and 
$u_0(L_i, R_i,X_i)=\sum_{k: L_i<t_k\leq R_i}\lambda_{k0}+\beta_0^\top \{Z_{x_i}(R_i)-Z_{x_i}(L_i)\}$. We now present the main result in the following theorem, whose proof is given in the Appendix. 
\begin{Th}\label{ourlemma1}
The minorization function for $\ell(\lambda,\beta)$ is
$\ell_{\dagger}(\lambda,\beta|\lambda_0,\beta_0)$, such that 
 $\ell(\lambda,\beta)\ge \ell_{\dagger}(\lambda,\beta|\lambda_0,\beta_0)$ $\forall \lambda,  \lambda_0>0$ and 
 $\beta, \beta_0\in \mathcal{R}^p$ and the equality holds when $\lambda=\lambda_0$ and $\beta=\beta_0$, and 
\bse
 \ell_{\dagger}(\lambda,\beta|\lambda_0,\beta_0)
\equiv \sum_{k=1}^m\mathcal{M}_{1,k}(\lambda_k|\lambda_0,\beta_0)+\mathcal{M}_2(\beta|\lambda_0,\beta_0)+\mathcal{M}_3(\lambda_0,\beta_0),
\ese
where 
\bse
&&\mathcal{M}_{1,k}(\lambda_k|\lambda_0,\beta_0)\\
&\equiv&
-\frac{\lambda_{k0}^2}{\lambda_k}   \sum^n_{i=1}\left\{ \frac{\Delta_{L,i}}{u_0(L_i,X_i)}I(t_k\leq L_i)
+
\frac{\Delta_{I,i}}{u_0(L_i,R_i,X_i)}I(L_i< t_k\leq R_i)\right\}\\ %
&& + \lambda_k
\sum_{i=1}^n\biggl[\Delta_{L,i}\left\{A_1(u_0(L_i,X_i))+2A_2(u_0(L_i,X_i))u_0(L_i,X_i)-\frac{1}{u_0(L_i,X_i)}\right\}I(t_k\leq L_i)\\
&&\hskip 10mm +\Delta_{I,i}\left\{A_1(u_0(L_i,R_i,X_i))+2A_2(u_0(L_i,R_i,X_i))u_0(L_i,R_i,X_i)
-\frac{1}{u_0(L_i,R_i,X_i)}\right\}\\
&& \hskip 10mm \times I(L_i< t_k\leq R_i)-\Delta_{I,i}I(t_k\leq L_i)-
\Delta_{R,i}I(t_k\leq R_i)\biggl]\\
&&
-\frac{\lambda_k^2}{\lambda_{k0}}
\sum^n_{i=1}\biggl\{\Delta_{L,i}A_2(u_0(L_i,X_i))u_0(L_i,X_i)I(t_k\leq L_i)\\
&&\hskip 10mm +\Delta_{I,i}A_2(u_0(L_i,R_i,X_i))u_0(L_i,R_i,X_i)I(L_i< t_k\leq R_i)\biggl\},\quad k=1,\ldots,m
\ese
\bse
&&\mathcal{M}_2(\beta|\lambda_0,\beta_0)\\
&\equiv&
-\sum_{i=1}^n\biggl[\frac{\Delta_{L,i}}{ u_0(L_i,X_i) }\times \frac{\{\beta_0^\top Z_{x_i}(L_i)\}^2}{
\beta^\top Z_{x_i}(L_i)}+\frac{\Delta_{I,i}}{u_0(L_i,R_i,X_i)   }\times 
\frac{\{\beta_0^\top (Z_{x_i}(R_i)-Z_{x_i}(L_i))\}^2}{\beta^\top (Z_{x_i}(R_i)-Z_{x_i}(L_i))}\biggl]\\
&&+
\sum_{i=1}^n\biggl[\Delta_{L,i}\left\{A_1(u_0(L_i,X_i))+2A_2(u_0(L_i,X_i))u_0(L_i,X_i)
-
\frac{1}{u_0(L_i,X_i)}\right\}\beta^\top Z_{x_i}(L_i)\\
&&\hskip 10mm +\Delta_{I,i}\left\{A_1(u_0(L_i,R_i,X_i))+2A_2(u_0(L_i,R_i,X_i))u_0(L_i,R_i,X_i)
- \frac{1}{u_0(L_i,R_i,X_i)}\right\}\\
&&\hskip 10mm \times \beta^\top \{Z_{x_i}(R_i)-Z_{x_i}(L_i)\}
-\Delta_{I,i}\beta^\top Z_{x_i}(L_i)-\Delta_{R,i}\beta^\top Z_{x_i}(R_i)\biggl]\\
&&-
\sum_{i=1}^n\biggl(\Delta_{L,i}A_2(u_0(L_i,X_i))\frac{u_0(L_i,X_i)}{\beta_0^\top Z_{x_i}(L_i)}\{\beta^\top Z_{x_i}(L_i)\}^2\\
&&\hskip 10mm +\Delta_{I,i}A_2(u_0(L_i,R_i,X_i))\left\{\frac{u_0(L_i,R_i,X_i)}{\beta_0^\top (Z_{x_i}(R_i)-Z_{x_i}(L_i))}\right\}[\beta^\top \{Z_{x_i}(R_i)-Z_{x_i}(L_i)\}]^2
\bigg),
\ese
$A_1(u)=\exp(-u)/\{1-\exp(-u)\}$, $A_2(u)=\exp(-u)/2\{1-\exp(-u)\}^2$
and the expression of $\mathcal{M}_3(\lambda_0,\beta_0)$ is given in the appendix. 
\end{Th}
As opposed to a direct maximization of $\ell(\lambda, \beta)$, for a given $(\lambda_0, \beta_0)$, the MM algorithm maximizes  $\ell_{\dagger}(\lambda,\beta|\lambda_0,\beta_0)$ with respect to $\lambda$ and $\beta$. In the next step, these new estimates replaces $(\lambda_0,\beta_0)$, followed by the maximization of $\ell_{\dagger}(\lambda,\beta|\lambda_0,\beta_0)$ with respect to $(\lambda,\beta)$. The iteration continues, until $(\lambda,\beta)$ and $(\lambda_0,\beta_0)$ are sufficiently close. It is important to note that 
although the MM and EM algorithms appear similar in their iterative way of function maximization, they differ in terms of the objective function that is being maximized. The paper by \cite{zhou2012vs} nicely articulates the similarities and differences between the EM and MM algorithms via a case study. In the EM algorithm, a conditional expectation of the complete data likelihood is maximized, whereas, in the MM, the minorization function of the log-likelihood is maximized. Most importantly, our specific choice of the minorization function allows separation of the parameters, thereby easing the maximization process. Furthermore, $\mathcal{M}_{1,k}(\lambda_k|\lambda_0, \beta_0)$ and $\mathcal{M}_2(\beta|\lambda_0, \beta_0)$ turned out to be concave functions of $\lambda_k$ and $\beta$ respectively.

To ensure the positivity of $\lambda_k,k=1,\ldots,m$, we use the  transformed parameters $\eta_k=\log(\lambda_k),k=1,\dots,m$ in the optimization. Define $\eta=(\eta_1,\ldots,\eta_m)^\top $ and $\eta_0=(\eta_{10},\ldots,\eta_{m0})^\top $, and  then replace 
$\lambda$ and $\lambda_0$ by $\exp(\eta)$ and  
$\exp(\eta_0)$, respectively, in $\mathcal{M}_{1, k}$ and $\mathcal{M}_2$ of the minorization function. Also, hereafter, we will refer to $\ell(\lambda, \beta)$ by $\ell(\eta, \beta)$. Consequently, the minorization function of $\ell(\eta, \beta)$ is $\ell_\dagger(\eta, \beta)$, obtained from $\ell_\dagger(\lambda, \beta)$ after replacing $\lambda$ and $\lambda_0$ by $\exp(\eta)$ and $\exp(\eta_0)$, respectively.

Next, we propose to estimate $\eta_k$ by solving $S_{1, k}(\eta_k|\eta_0,\beta_0)  \equiv\partial \mathcal{M}_{1,k}(\exp(\eta_k)|\exp(\eta_0),\beta_0)/\partial$ $\eta_k=0$  for $k=1,\ldots,m$ and $\beta$ by solving  $S_2(\beta|\eta_0,\beta_0)\equiv\partial \mathcal{M}_2(\beta|\exp(\eta_0),\beta_0)/\partial\beta=0$. Note that given $(\eta_0, \beta_0)$, $S_{1, k}(\eta_k|\eta_0, \beta_0)$ is 
a function of only the scalar parameter $\eta_k$. Now, following the general strategy of gradient MM algorithm \citep{Hunter2004}, given $(\eta_0, \beta_0)$, $(\eta, \beta)$ will be updated by one step Newton-Raphson method, and the entire method can be summarized in the following steps.

\vskip 3mm 

\noindent    
Step 0. Initialize $(\eta, \beta)$. 

\noindent 
Step 1. At the $\iota$th step of the iteration, we update the parameters as follows:
\begin{eqnarray}
\eta_k^{(\iota)}&=&\eta_k^{(\iota-1)}-S^{-1}_{1, kk}(\eta_k^{(\iota-1)}|\eta^{(\iota-1)},\beta^{(\iota-1)})S_{1, k}(\eta_k^{(\iota-1)}|\eta^{(\iota-1)},\beta^{(\iota-1)}), \mbox{ for } k=1,\ldots,m,\label{eq:mmeta}\\
\beta^{(\iota)}&=&\beta^{(\iota-1)}-S^{-1}_{22}(\beta^{(\iota-1)}|\eta^{(\iota-1)},\beta^{(\iota-1)})S_2(\beta^{(\iota-1)}|\eta^{(\iota-1)},\beta^{(\iota-1)}), \label{eq:mmbeta}
\end{eqnarray}
where $(\eta^{(\iota-1)}, \beta^{(\iota-1)})$ and 
$(\eta^{(\iota)}, \beta^{(\iota)})$ denote the parameter estimates at the $(\iota-1)$th and $\iota$th iterations, respectively. 

\noindent 
Step 3. Repeat Step 1 until $(\eta^{(\iota-1)}, \beta^{(\iota-1)})$ and 
$(\eta^{(\iota)}, \beta^{(\iota)})$ are sufficiently close. 

\vskip 3mm 

In the above iteration both $S_{1, k}$ and $S_{1, kk}$ are scalar valued functions, and  $S_2$ is a $p$-dimensional vector while $S_{22}$ is a $p\times p$ matrix. After the convergence, the final estimate of $\beta$ and $\eta$ will be denoted by $\wh\beta$ and $\wh\eta$. 
The expression of the terms involved in 
(\ref{eq:mmeta}) and (\ref{eq:mmbeta}) 
are 
\begin{eqnarray}
&&S_{1,k}(\eta_k^{\iota-1}|\eta^{\iota-1},\beta^{\iota-1}) \nonumber\\
&=&\exp(\eta_k^{\iota-1})\sum_{i=1}^n\biggl\{\Delta_{L,i}A_1(u_{(\iota-1)}(L_i,X_i))I(t_k\leq L_i)-\Delta_{I,i}I(t_k\leq L_i)-\Delta_{R,i}I(t_k\leq R_i)\nonumber\\
&&\quad\quad\quad\quad\quad+\Delta_{I,i}A_1(u_{(\iota-1)}(L_i,R_i,X_i))I(L_i< t_k\leq R_i)\biggl\},\quad k=1,\ldots,m,\label{eqs1k}\\
&&S_{1,kk}(\eta_k^{\iota-1}|\eta^{\iota-1},\beta^{\iota-1})\nonumber\\
&=&\exp(\eta_k^{\iota-1})\sum_{i=1}^n\biggl[\Delta_{L,i}\Bigg\{A_1(u_{(\iota-1)}(L_i,X_i))-2A_2(u_{(\iota-1)}(L_i,X_i))u_{(\iota-1)}(L_i,X_i)\nonumber\\
&&\quad\quad\quad\quad\quad
-\frac{2}{u_{(\iota-1)}(L_i,X_i)}
\Bigg\}I(t_k\leq L_i) -\Delta_{I,i}I(t_k\leq L_i)-\Delta_{R,i}I(t_k\leq R_i)\nonumber\\
&&\quad\quad\quad\quad\quad+\Delta_{I,i}\Bigg\{A_1(u_{(\iota-1)}(L_i,R_i,X_i))-2A_2(u_{(\iota-1)}(L_i,R_i,X_i))u_{(\iota-1)}(L_i,R_i,X_i)\nonumber\\
&&\quad\quad\quad\quad\quad\quad\quad\quad-\frac{2}{u_{(\iota-1)}(L_i,R_i,X_i)}
\Bigg\}I(L_i< t_k\leq R_i)\biggl],\quad k=1,\ldots,m,\label{eqs1kk}\\
&&S_2(\beta^{(\iota-1)}|\eta^{(\iota-1)},\beta^{(\iota-1)})\nonumber\\
&=&\sum_{i=1}^n\biggl\{\Delta_{L,i}A_1(u_{(\iota-1)}(L_i,X_i))Z_{x_i}(L_i)-\Delta_{I,i}Z_{x_i}(L_i)-\Delta_{R,i}Z_{x_i}(R_i)\nonumber\\
&&\quad +\Delta_{I,i}A_1(u_{(\iota-1)}(L_i,R_i,X_i))(Z_{x_i}(R_i)-Z_{x_i}(L_i))\biggl\},\nonumber\\
&&S_{22}(\beta^{(\iota-1)}|\eta^{(\iota-1)},\beta^{(\iota-1)})\nonumber\\
&=&-2\sum_{i=1}^n\biggl[\Delta_{L,i}\Bigg\{A_2(u_{(\iota-1)}(L_i,X_i))u_{(\iota-1)}(L_i,X_i)+\frac{1}{u_{(\iota-1)}(L_i,X_i)}\Bigg\}\frac{Z_{x_i}(L_i)^{\otimes 2}}{Z_{x_i}(L_i)^\top \beta^{(\iota-1)}}\nonumber\\
&&\quad\quad 
+\Delta_{I,i}\Bigg\{A_2(u_{(\iota-1)}(L_i,R_i,X_i))u_{(\iota-1)}(L_i,R_i,X_i)+\frac{1}{u_{(\iota-1)}(L_i,R_i,X_i)}\Bigg\}\nonumber\\
&&\quad\quad \times \frac{(Z_{x_i}(R_i)-Z_{x_i}(L_i))^{\otimes 2}}{(Z_{x_i}(R_i)-Z_{x_i}(L_i))^\top \beta^{(\iota-1)}}\biggl],\nonumber
\end{eqnarray}
where $u_{\iota-1}(L_i, X_i)$, $u_{\iota-1}(R_i, X_i)$ and $u_{\iota-1}(L_i, R_i,  X_i)$ are the  $u_{0}(L_i, X_i)$, $u_{0}(R_i, X_i)$ and $u_{0}(L_i, R_i,  X_i)$,
 with $\beta_0$ and $\lambda_0$ replaced by 
 $\beta^{(\iota-1)}$ and $\exp(\eta^{(\iota-1)})$, respectively. 
 For the computation of the estimator or the standard error, if any term (expression) turns out to be $0/0$, it is re-defined as $0$.

\subsection{Variance estimation}
\cite{Zeng2006} studied the asymptotic properties of the ML estimator, and used the profile likelihood method \citep{Murphy2000} to calculate the asymptotic standard error of the estimator. We also follow their idea of the standard error calculation, which will be aided by our computational tools. 
Specifically, the authors studied consistency of the estimator of 
$\beta$ and $\Lambda(t)=\int^t_0 \lambda(u)du$, the baseline cumulative  hazard function, and the asymptotic property of $\widehat\beta$. 
Suppose that the estimator of the covariance matrix of $\wh\beta$ is $-D^{-1}$. Then, the $(r, s)$th element of the $p\times p$ matrix $D$ is
\bse
\frac{{\rm pl}(\wh{\beta})-{\rm pl}(\wh{\beta}+h_ne_r)-{\rm pl}(\wh{\beta}+h_ne_s)+{\rm pl}(\wh{\beta}+h_ne_r+h_ne_s)}{h_n^2},
\ese
with $e_r$ being  the $p\times 1$ vector with 1 at the $r$th position and 0 elsewhere, $h_n$ is a constant with an order $n^{-1/2}$, and ${\rm pl}(\beta)$ stands for the profile log-likelihood function  defined as 
${\rm pl}(\beta)=\ell(\wh\eta^\beta,\beta)$, 
where  $\wh{\eta}^{\beta}={\rm argmax}_{\eta\in\mathcal{R}^m}\ell(\eta,\beta)$. To obtain  $\wh{\eta}^{\beta}$, we  use the proposed minorization function, and specifically use the $m$ equations given in 
 (\ref{eq:mmeta}) after replacing 
 $\beta^{\iota-1}$ to $\beta$.

Specifically, to obtain $\wh{\eta}^{\beta}$, we shall  maximize the log-likelihood function $\ell(\eta,\beta)$  with respect to $\eta$ only. The minorization function for $\ell(\lambda,\beta)$ is $\ell_\dagger(\lambda,\beta|\lambda_0,\beta_0=\beta)$. Since $\beta$ is fixed, we only need to maximize functions $\mathcal{M}_{1,k}(\lambda_k|\lambda_0,\beta)$ for $k=1,\ldots,m$. Following the general strategy of gradient MM algorithm, at the $\iota$th step of the iteration, $\eta_k^{(\iota)}(=\log(\lambda^{(\iota)}_k))
$ is updated as follows,
 \bse
 \eta_k^{(\iota)}&=&\eta_k^{(\iota-1)}-S^{-1}_{1, kk}(\eta_k^{(\iota-1)}|\eta^{(\iota-1)},\beta)S_{1, k}(\eta_k^{(\iota-1)}|\eta^{(\iota-1)},\beta), \mbox{ for } k=1,\ldots,m,
 \ese
where $S_{1, k}(\eta_k^{(\iota-1)}|\eta^{(\iota-1)},\beta)$ and $S_{1, kk}(\eta_k^{(\iota-1)}|\eta^{(\iota-1)},\beta)$ are $S_{1, k}(\eta_k^{(\iota-1)}|\eta^{(\iota-1)},\beta^{(\iota-1)})$ and $S_{1, kk}(\eta_k^{(\iota-1)}$    $|\eta^{(\iota-1)},\beta^{(\iota-1)})$, respectively, when $\beta^{(\iota-1)}$ is set to $\beta$. The expression of $S_{1, k}(\eta_k^{(\iota-1)}|\eta^{(\iota-1)},\beta^{(\iota-1)})$
and $S_{1, kk}(\eta_k^{(\iota-1)}|\eta^{(\iota-1)},\beta^{(\iota-1)})$ are given in (\ref{eqs1k}) and (\ref{eqs1kk}), respectively.

For any given $\beta$, the computation of $\wh{\eta}^{\beta}$  is very fast when 
$\wh\eta=(\wh\eta_1,\ldots,\wh\eta_m)^\top $, the MLE,  is used as the initial value. Obtaining $\wh{\eta}^{\beta}$ 
using any generic optimization of $\ell(\eta, \beta)$ can be very time consuming.

\subsection{Complexity analysis}\label{sec:complexity}
In the proposed method, parameters are updated via equations (\ref{eq:mmeta}) and (\ref{eq:mmbeta}). Now, we inspect the computational complexity (or simply complexity) of a single update.   
 The  complexity to calculate $S_2(\beta|\eta,\beta)$ and $S_{22}(\beta|\eta,\beta)$ is $O(np+np^2)$, where $n$ is the sample size. Next, the  complexity of inverting $S_{22}(\beta|\eta,\beta)$ is $O(p^3)$. Therefore, the complexity of one update of $\beta$ is $O(np+np^2+p^3)$. Similarly, for any $k=1, \dots, m$,  the complexity of one step update of $\eta_k$ is $O(2n+1)$. Hence, the total computational cost for updating $\eta$ and $\beta$ is $O((2n+1)m+np+np^2+p^3)$.

Now, we look closely the computational complexity of the generic optimization of the log-likelihood $\ell(\lambda, \beta)$ (aka $\ell(\exp(\eta), \beta)$) using the Newton-Raphson approach.  In each step, the computational cost of gradient and the Hessian matrix of the log-likelihood is $O(n(m+p)+n(m+p)^2)$, and inverting a matrix of order $m+p$ will cost $O((p+m)^3)$. The total complexity for a single update  is then  $O(n(p+m)+n(m+p)^2+(p+m)^3)$, which is obviously larger than $O((2n+1)m+np+np^2+p^3)$. Since $m$ increases with the sample size $n$, the difference between the two complexities increases with $n$. Alternative to Newton's method, if the Broyden–Fletcher–Goldfarb–Shanno (BFGS) algorithm \citep{fletcher2013practical} is used, the complexity becomes 
$O(n(m+p)+(n+1)(m+p)^2)$. Note the BFGS algorithm avoids matrix inversion, so the cubic order complexity is avoided. The complexity of the BFGS method involves $m^2$ and $p^2$ term, whereas the complexity of the proposed method has $m$ and $p^3$ term. Usually, 
for the semiparametric regression model, $p$ is much smaller than $m$
that tends to increase with $n$, indicating the complexity of MM is smaller than BFGS in this context. This complexity calculation indicates the advantage of the MM algorithm.

\section{Simulation study}\label{sec:simulation}
In this section, we conducted a numerical study to assess the finite-sample performances of the proposed MM algorithm. We considered two main scenarios, 1) time-independent  and 2) time-dependent covariates. 
For Scenario 1, we simulated a scalar covariate $X$  from 
${\rm Bernoulli}(0.5)$. Conditional on the covariate, we considered the following  hazard function
$
h(t|X)=0.2+\beta X$. 
For Scenario 2, the hazard function was 
$
h(t|X)=0.2+\beta X\exp(t), 
$ with $X\sim {\rm Bernoulli}(0.5)$. We considered two different values of $\beta$, 0.5 and 1. 
For both scenarios,  we simulated the left censoring time $L_i$  from ${\rm Uniform}(0.1,\,   2)$ and the right censoring time $R_i$  from  ${\rm Uniform}(L_i+0.5,  4)$. 
The proportion of left censoring was from 30\% to 50\% and the proportion of right censoring was 
from 25\% to 35\% across all the scenarios. For each scenario, we considered three sample sizes, $n=100$, $200$ and  $500$.  For the profile likelihood based standard error calculation, we used $h_n=1.5n^{-1/2}$ because among several trial values of $h_n$ this one yielded good agreement between the standard deviation and the standard error of the estimators. We have not faced any convergence issue in our  proposed MM algorithm. 

We fit the ARM (\ref{eqm1}) to each of the simulated dataset using the proposed MM algorithm. The results of the simulation study with $500$ replications are presented in Table \ref{simumytab1}. 
\begin{table}[h]
\begin{center}
\begin{threeparttable}
\addtolength{\tabcolsep}{-4pt}
\caption{Results of the simulation study with a scalar covariate, for both time-independent and time-dependent scenarios. Est: the average of the estimates, SD: the standard deviation of the estimates, SE: the average of the standard errors, CP: the coverage probability of the 95\% Wald's confidence interval} \label{simumytab1}
	\vspace{0.01 in}
	{\small
		\begin{tabular}{p{1cm}p{2cm}             p{1cm} p{1cm}p{1cm}p{1cm}p{1cm}p{1cm}p{1cm}p{1cm}p{1cm} p{1cm}p{1cm}p{1cm}p{1cm} }
			\hline
			\multicolumn{14}{c}{Time-independent covariate: $h(t|X)=0.2+\beta X$}\\
			& &  \multicolumn{4}{c}{$n=100$} &  \multicolumn{4}{c}{$n=200$}&\multicolumn{4}{c}{$n=500$}\\
		 $\lambda(t)$& $\beta$ & Est &SD &SE &CP& Est &SD &SE &CP & Est &SD &SE &CP\\
		 \hline
		 0.2 & $0.5$&$0.495$ & $0.145$ & $0.150$ & $0.956$ & $0.496$ & $0.096$ & $0.099$ & $0.952$ & $0.499$ & $0.059$ & $0.058$ & $0.946$\\
		 0.2 & $1.0$&$1.047$ & $0.222$ & $0.248$ & $0.978$ & $1.005$ & $0.161$ & $0.160$ & $0.944$ & $1.012$ & $0.100$ & $0.091$ & $0.936$\\
		           \hline \multicolumn{14}{c}{Time-dependent covariate: $h(t|X)=0.2+\beta X\exp(t)$}\\
            	& &  \multicolumn{4}{c}{$n=100$} &  \multicolumn{4}{c}{$n=200$}&\multicolumn{4}{c}{$n=500$}\\
		 $\lambda(t)$& $\beta$ & Est &SD &SE &CP& Est &SD &SE &CP & Est &SD &SE &CP\\
		 \hline
		 0.2 & $0.5$&$0.518$ & $0.134$ & $0.160$ & $0.992$ & $0.504$ & $0.090$ & $0.102$ & $0.980$ & $0.505$ & $0.053$ & $0.059$ & $0.974$\\
		 0.2 & $1.0$&$1.085$ & $0.314$ & $0.317$ & $0.986$ & $1.040$ & $0.200$ & $0.202$ & $0.978$ & $1.013$ & $0.110$ & $0.113$ & $0.950$\\
    		 \hline
					\end{tabular}

		}
	\end{threeparttable}
	\end{center}
\end{table}
For each scenario, we report the average of the estimates (Est) for $\beta$, empirical standard deviation (SD), the average of the estimated standard error (SE), and the 95\% coverage probability (CP) based on Wald's confidence interval. The results indicate that the  proposed MM algorithm can estimate the parameters very well, while the bias could be up to $8.5\%$ across all scenarios. Overall, the bias and SD decrease with the sample size $n$. There is a reasonable agreement between the empirical standard deviation and the estimated standard error. The CPs are pretty close to the nominal level, $0.95$.

To assess the performance of the algorithm for the multiple covariates scenario, we conducted another simulation study with 
$h(t|X_1,X_2)=0.2t^{1/2}+\beta_1 X_1+\beta_2 X_2$. We simulated both covariates $X_1$ and $X_2$ from 
from Bernoulli(0.5), and set  $\beta_1=0.5$ and  $\beta_2=1$. After simulating the time-to-event $T$ using the additive hazard $h(t|X_1,X_2)$, the we simulated the left-censoring time $L$ from Uniform(0.1,\, 1.5) and the right-censoring time $R$  from ${\rm Uniform}(L+1.5,\, 4)$.
This resulted in  42\%  left censored, 42\% interval censored, and 16\% right  censored subjects. We fit ARM  (\ref{eqm1}) to each of the simulated datasets. 
  We observe the adequate performance of our proposed algorithm (Table \ref{simumytab2}), with results similar to Table \ref{simumytab1}. 

\begin{table}[h]
\begin{center}
\begin{threeparttable}
\addtolength{\tabcolsep}{-4pt}
\caption{Results of the simulation study with two covariates, $X_1\sim {\rm Bernoulli(0.5)}$ and $X_2\sim {\rm Bernoulli(0.5)}$. Est: the average of the estimates, SD: the standard deviation of the estimates, SE: the average of the standard errors, CP: the coverage probability of the 95\% Wald's confidence interval} \label{simumytab2}
	\vspace{0.01 in}
	{\small
		\begin{tabular}{p{2cm}             p{1cm} p{1cm}p{1cm}p{1cm}p{1cm}p{1cm}p{1cm}p{1cm}p{1cm} p{1cm}p{1cm}p{1cm}p{1cm} }
		\hline
          	 &  \multicolumn{4}{c}{$n=100$} &  \multicolumn{4}{c}{$n=200$}&\multicolumn{4}{c}{$n=500$}\\
		   & Est &SD &SE &CP& Est &SD &SE &CP & Est &SD &SE &CP\\
		   \hline
		  $\beta_1=0.5$&$0.490$ & $0.193$ & $0.202$ & $0.958$ & $0.493$ & $0.127$ & $0.130$ & $0.950$ & $0.501$ & $0.077$ & $0.076$ & $0.940$\\
		 $\beta_2=1.0$&$1.027$ & $0.287$ & $0.287$ & $0.968$ & $1.021$ & $0.181$ & $0.186$ & $0.964$ & $1.010$ & $0.107$ & $0.104$ & $0.934$ \\
\hline
					\end{tabular}

		}
	\end{threeparttable}
	\end{center}
\end{table}

In all computations, the iteration is stopped when the sum of the absolute differences of the estimates for $\eta$ and $\beta$ at two successive iterations is less than $10^{-3}$. All computations were conducted in an Intel(R) Xeon(R) CPU E5-2680 v4 at 2.40 GHz machine.
In Table \ref{tabcom}, we provide the average computation times to obtain parameter estimates and the standard errors for varying sample sizes and  the scalar covariate and the two covariates scenarios using the proposed method and the direct optimization of the log-likelihood using the BFGS algorithm. Here, the specific form of log-likelihood function is given in the expression \eqref{log-likelihood}. To derive estimates using the BFGS algorithm, we first coded the negative of the log-likelihood function and used it as one of the input arguments of the \texttt{optim} function in \texttt{R} with the BFGS method. The initial values were the same as that in the proposed MM algorithm. The standard errors of the estimates are the square root of the diagonal of the inverse of the negative Hessian matrix  which is returned from the optimization.
\begin{table}[h]
\begin{center}
\caption{The average time (in seconds) to compute estimates (ATE) and 
standard errors (ATS). Case 1: scalar covariate; Case 2: two covariates; MM: proposed MM algorithm; Direct: direct optimization} \label{tabcom}
	\vspace{0.02 in}
		\begin{tabular}{ll  rr rr rr} 
		\hline
      &         &  \multicolumn{2}{c}{$n=100$} &  \multicolumn{2}{c}{$n=200$}&\multicolumn{2}{c}{$n=500$}\\
	  &	        & ATE &ATS     & ATE &ATS   & ATE &ATS  \\
	  \hline
 Case 1& MM     & 1.08      & 0.39     &     11.92  &   7.33    &  78.96    &  80.04  \\    
       & Direct &  3.50     &  1.24    &    37.79   & 18.88      & 1587.08      &666.62 \\
 Case 2& MM     &  1.91 & 1.88 & 13.14 & 16.93 & 87.78 & 208.13\\
       & Direct &  8.32 & 6.23 & 92.81 & 65.10 & 1988.76 & 1812.97 \\
\hline
\end{tabular}
	\end{center}
\end{table}

The results show that the proposed method is several times faster than the direct optimization of the log-likelihood function. The relative gain in the computation time increases with the sample size.

\section{Application: Breast Cancer Data}\label{sec:realdata}To illustrate the proposed method, we analyzed the breast cancer data considered in \cite{Finkelstein1986} and \cite{Finkelstein1985}. In this breast cosmesis study, the subjects under the adjuvant chemotherapy after tumorectomy were periodically followed-up for the cosmetic effect of the therapy. So, patients generally visited the clinic every 4 to 6 months. Thus, the time of the appearance of breast retraction was recorded as an interval. In particular, if the recorded time for a patient is $(0, 4]$, then the breast retraction happened before four months, whereas, if for any subject the time to the occurrence is $(6, 12]$, then it signifies that the event had happened between six and twelve months. There were 94 early breast cancer patients in the study, of which 46 patients were given radiation therapy alone, and 48 patients were given radiation therapy plus adjuvant chemotherapy. The analysis aimed to study the effect of chemotherapy on time until the appearance of retraction. 

\begin{figure}[p]
\centering
  \includegraphics[width = \textwidth, height = 12cm]{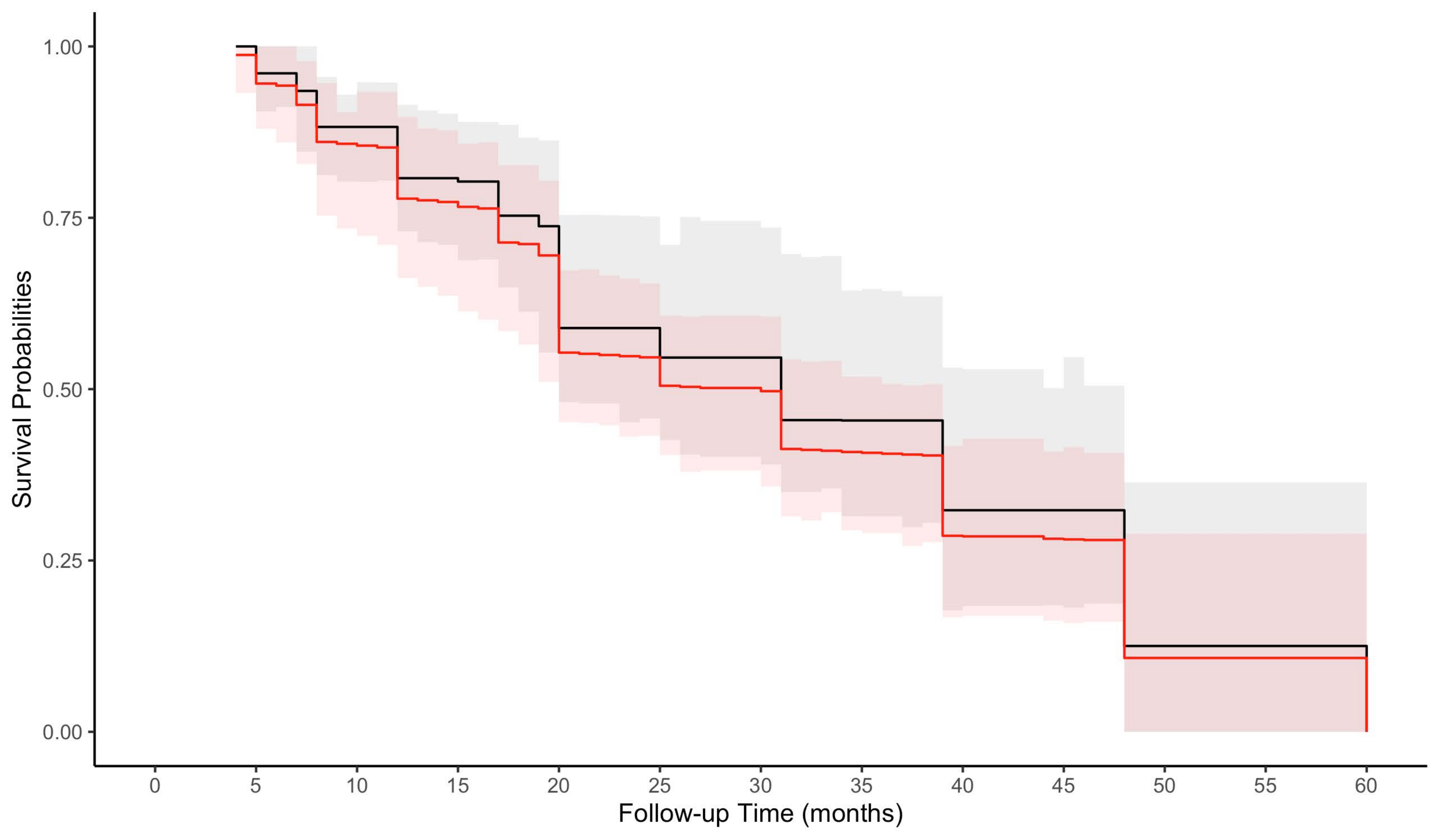}
\vspace*{-10mm}  
\caption{Estimated survival curves of the breast cancer data. The red and black curves correspond to patients with  $X=1$ (adjuvant chemotherapy $+$ radiation) \label{fig:SurvivalCurves}
and $X=0$ (only radiation), respectively. The pink and gray shaded areas are the confidence bands for red and black curves, respectively.}  
\end{figure}

We set $X=1$ if a patient had received adjuvant chemotherapy following the initial radiation treatment and 0 otherwise. Hence, $X$ is a time independent covariate, and we fit the model $h(t|X)=\lambda(t)+X\beta$ to the data using the proposed method. Here, $\beta$ represents the difference in the hazard of breast retraction between $X=1$ and $X=0$ groups at any time point. We  obtain $\wh{\beta}=0.031$. Since the choice of $h_n$ was quite arbitrary in the profile likelihood-based method of standard error, we used different values of $h_n$,  $1.5n^{-1/2}$, $n^{-1/2}/20$, $n^{-1/2}/100$ and $n^{-1/2}/1000$, and obtained $0.09$, $0.08$, $0.06$ and $0.007$ as the standard errors.  Obviously, for standard error $0.007$, $\wh\beta$ is significantly different from zero at the $5\%$ level, while for other standard errors $\wh\beta$ is not significantly different from zero. To investigate this issue further, we calculated bootstrap standard errors using $200$ bootstrap samples, which came out to be 0.06. Figure \ref{fig:SurvivalCurves} plots the estimated survival curves for the two groups along with their 95\% pointwise confidence intervals calculated using the bootstrap method. This analysis shows no significant difference between the two survival functions or the two hazards functions at any time. On the contrary, \cite{Finkelstein1986} fit a proportional hazard model to this data and found a statistically significant effect of chemotherapy. 
  \begin{center}
  {\bf [Figure 1 should be here]}    
  \end{center}

\section{Implementation: \texttt{R} package \texttt{MMIntAdd}}\label{sec:forR} 
For the implementation of our proposed method, we have developed an \texttt{R} package, and it is available at GitHub: \url{https://github.com/laozaoer/MMIntAdd}. In this section, we discuss how the package can be used to analyze the breast cosmesis dataset. The first step is installing the package. One can use the \texttt{R} package  \texttt{devtools} to install our \texttt{R} package as follows. 
\begin{lstlisting}[language=R]
>library(devtools)
>devtools::install_github("laozaoer/MMIntAdd")
\end{lstlisting}
If the above method fails, then alternatively one may use the \texttt{remotes} package to install \texttt{MMIntAdd}. The code is 
\begin{lstlisting}[language=R]
>library(remotes)
>remotes::install_github("laozaoer/MMIntAdd")
\end{lstlisting}
During the installation, when asked, it is customary to update the dependent packages, \texttt{Rcpp}, 
\texttt{RcppArmadillo}, or \texttt{boot}. 
After installation, load the package in the \texttt{R} console using the command
\begin{lstlisting}[language=R]
>library(MMIntAdd)
\end{lstlisting}
Let us now analyze the breast cosmesis data available in the package. This dataset was taken from the \texttt{interval} package and reformatted. Unlike the description 
 given in Section 2,  the first two columns of the dataset do not represent the finite 
 inspection time window; rather, they represent the two boundary points of the time-to-event. Specifically, for a left-censored subject, the entry in the first column is zero, while the entry of the second column is infinity for a right-censored subject. The following three columns are left-, interval-, and right-censoring indicators. Note that the sum of these indicators must be equal to one for any subject. The sixth column of the data represents the covariate value. 
\begin{lstlisting}[language=R]
> data(bcos)
> head(bcos)
  left right L I R covariate
1   45   Inf 0 0 1         0
2    6    10 0 1 0         0
3    0     7 1 0 0         0
4   46   Inf 0 0 1         0
5   46   Inf 0 0 1         0
6    7    16 0 1 0         0
\end{lstlisting}
There are two functions of the \texttt{MMIntAdd} package, \verb=Add_case2_inte= and \verb=Add_ci_boot=.  To find them, use the command
\begin{lstlisting}[language=R]
> lsf.str("package:MMIntAdd")
Add_case2_inte : function (data, hn.m, Max_iter = 1000, Tol = 0.001)
Add_ci_boot : function (data, time_points, covariate_value, CItype =
c("norm", "basic","perc", "bca"), conf = 0.95, boot.num = 200, 
object_type =c("reg"), Max_iter = 1000, Tol = 0.001)
\end{lstlisting}
The first function returns the regression parameter estimates and the standard error calculated using the profile likelihood approach. For the standard error calculation, we require the bandwidth that is given as an input argument, hn.m of the function. Different values of hn.m returns different standard errors but with the same parameter estimates. 
\begin{lstlisting}[language=R]
> result_hn1=Add_case2_inte(bcos,hn.m=1.5)
> print(result_hn1$beta)
            Est         SE
[1,] 0.03136608 0.09057521
> result_hn2=Add_case2_inte(bcos,hn.m=1/20)
> print(result_hn2$beta)
            Est         SE
[1,] 0.03136608 0.08259436
> result_hn3=Add_case2_inte(bcos,hn.m=1/100)
> print(result_hn3$beta)
            Est         SE
[1,] 0.03136608 0.05657365
> result_hn4=Add_case2_inte(bcos,hn.m=1/1000)
> print(result_hn4$beta)
            Est       SE
[1,] 0.03136608 0.007612
\end{lstlisting}
The other returned objects of \verb=Add_case2_inte= are the estimates of $\lambda=(\lambda_1,\ldots,\lambda_m)^\top $,  the log-likelihood value and the set of distinct inspection time points.

The other function of the \texttt{MMIntAdd} package is used to obtain the bootstrap standard error and confidence interval. There are many input arguments to that function. Among them, \texttt{boot.num} denotes the number of bootstrap samples to be used. 
\begin{lstlisting}[language=R]
> Add_ci_boot(bcos,boot.num = 200)
$beta_boot_se
                 Est    boot_se
covariate 0.03136608 0.06354992
$CI_beta
$CI_beta$normal
       index method        lwr       upr
normal     1 normal -0.1092515 0.1398596
$CI_beta$basic
      index method        lwr        upr
basic     1  basic -0.1277781 0.06273215
$CI_beta$percent
        index  method          lwr       upr
percent     1 percent 3.127308e-55 0.1905102
$CI_beta$bca
    index method          lwr      upr
bca     1    bca 3.142232e-37 0.240083
\end{lstlisting}

The above function returns bootstrap standard error and bootstrap confidence intervals of the regression parameter, which varies according to the method chosen. Although the default confidence level is 0.95, the level can be set to a different value. These functions can also handle multiple covariates. All the covariates must be binary or numeric, and they are placed from the sixth column onwards in the data frame. For analyzing data with a categorical covariate with $k$ nominal categories, the $(k-1)$  dummy variables must be incorporated in the data frame. 

Next, we analyze a simulated dataset using the \texttt{MMIntAdd} package. 

\begin{lstlisting}[language=R]
> set.seed(10)
> n=100
> # Generation of three covariates
> x1=rbinom(n, 1, 0.5) # the first covariate
> x2=rbinom(n, 1, 0.4) # the second covariate
> x3=rbinom(n, 1, 0.3) # the third covariate
> 
> #caplambda=0.2*t+ t*(0.5*x1+1*x2+0.6*x3), the true value of the
> #regression parameters are 0.5, 1 and 0.6.
> r=runif(n, 0, 1)
> time_to_event=-log(r)/(0.2+ 0.5*x1+1*x2+0.6*x3)
> # Generation of inspection time window (L, R)
> myl= runif(n,0.1,1.5)
> myr=runif(n, myl+1.5, 4)
> #### Censoring indicator
> delta_ell=as.numeric(time_to_event<myl)
> delta_r=as.numeric(time_to_event>myr)
> delta_i=1-delta_ell-delta_r
> 
> myr[delta_ell==1]=myl[delta_ell==1]
> myl[delta_ell==1]=0
> myl[delta_r==1]=myr[delta_r==1]
> myr[delta_r==1]=Inf
> # Creation of the final data object
> mydata=data.frame(myl, myr, delta_ell, delta_i, delta_r, x1,x2,x3)
> mydata=as.matrix(mydata)
> # Analysis of the data by invoking the following function
> testresult=Add_case2_inte(mydata,hn.m=1.5)
> testresult$beta
         Est        SE
x1 0.7008246 0.2899771
x2 1.0521943 0.3808892
x3 0.4904499 0.2564767
\end{lstlisting}

Suppose that, for this example, we are interested in obtaining the bootstrap standard error of the regression parameters and the bootstrap confidence interval of the survival probability at select time points and for a given set of covariate values. For illustration, suppose that the interest is in the survival probability at only two time points, 0.5 and 0.6, and for a covariate value of (0, 1, 0).  The code is 

\begin{lstlisting}[language=R]
> mytimepoints=c(0.5, 0.6)
> mycov=c(0, 1, 0)
> out=Add_ci_boot(mydata,time_points=mytimepoints,
+ covariate_value = mycov, object_type = c("reg","surv"))
> names(out)
[1] "beta_boot_se" "CI_beta"      "surv_boot_se" "CI_surv"     
> out$beta_boot_se
         Est   boot_se
x1 0.7008246 0.2365580
x2 1.0521943 0.3334506
x3 0.4904499 0.2776679
> out$CI_beta
$normal
        index method         lwr      upr
normal      1 normal  0.29612686 1.223417
normal1     2 normal  0.37157108 1.678673
normal2     3 normal -0.04340368 1.045034
$basic
       index method        lwr      upr
basic      1  basic  0.2265320 1.248173
basic1     2  basic  0.2608937 1.607011
basic2     3  basic -0.1692522 0.980646
$percent
         index  method          lwr      upr
percent      1 percent 0.1534763545 1.175117
percent1     2 percent 0.4973772979 1.843495
percent2     3 percent 0.0002538463 1.150152
$bca
     index method         lwr      upr
bca      1    bca 0.406808700 1.703756
bca1     2    bca 0.478899213 1.833237
bca2     3    bca 0.001413554 1.175776
> out$surv_boot_se
        Est   boot_se
1 0.4096685 0.1283081
2 0.2226586 0.1198392
> out$CI_surv
$normal
        index method        lwr       upr
normal      1 normal 0.23085708 1.0000000
normal1     2 normal 0.07454217 0.5108168
$basic
       index method        lwr       upr
basic      1  basic 0.27802103 1.0000000
basic1     2  basic 0.08806235 0.6798638
$percent
         index  method        lwr       upr
percent      1 percent 0.11100018 0.6036531
percent1     2 percent 0.07292174 0.5629745
$bca
     index method        lwr       upr
bca      1    bca 0.23570171 0.6307615
bca1     2    bca 0.02607938 0.4095530
\end{lstlisting}
After examining all the results, we recommend using the BCA confidence interval \citep{Efron1993} for the regression parameters and the survival probabilities.

\section{Conclusions}\label{sec:conclusion}
This chapter proposed an efficient MM algorithm to obtain ML estimates of a complex likelihood function for the ARM with interval-censored responses. The attractive feature of the method is enabling the separation of the finite and infinite dimensional parameters. This separation of components provides significant computational advantages as the dimension of the infinite-dimensional parameter increases with the sample size. Numerical studies show that the algorithm works well; we have not encountered any convergence issues in the simulation settings or real data analysis.

We believe that this MM proposal will help generate new ideas for handling computational bottlenecks in complex models and likelihoods. Model (\ref{eqm1}) assumes a constant effect of the covariate. However, rather than a constant regression parameter, one can consider a time-dependent coefficient $\beta(t)$ without specifying any form \citep{Huffer1991}. Some other interesting topics for future research include developing MM-based computationally efficient methods and algorithms for the clustered case-I or case-II interval-censored responses \citep{Huang1996, TongWang2020}, including exploration of big-data scalability in tune to recent advances via asynchronous distributed EM algorithms \citep{srivastava2019asynchronous}. Additionally, developing computationally efficient methods when the inspection time is informative \citep{Zhao2021} could also be a  direction of future research.

\section*{Acknowledgement}
Bandyopadhyay acknowledges funding support from the NIH/NCI grants P20CA252717, P20CA264067, and P30CA016059 (VCU's Massey Cancer Center Support Grant). 

\bibliographystyle{biom}
\bibliography{references}

\section*{Appendix}
\setcounter{equation}{0} 
\renewcommand{\theequation}{A.\arabic{equation}}
\setcounter{section}{0} 
\renewcommand{\thesection}{A.\arabic{section}}

We shall use  the second part  of Lemma 1 from \cite{TongWang2020} in proving Theorem 1, and we present this result  
in the following proposition.  
The proof of proposition \ref{prop1} can be found in \cite{TongWang2020}. 

\begin{proposition}\label{prop1}
%
%
\citep{TongWang2020} 
 For any $\tau,\tau_0\ge 0$
\bse
\log\left\{\frac{1-\exp(-\tau)}{1-\exp(-\tau_0)}\right\}\geq (\tau-\tau_0)A_1(\tau_0)-(\tau-\tau_0)^2A_2(\tau_0)+\log\left(\frac{\tau_0}{\tau}\right)+1-\frac{\tau_0}{\tau},
\ese
where $
A_1(\tau_0)=\exp(-\tau_0)/\{1-\exp(-\tau_0)\}$ 
and $A_2(\tau_0)=\exp(-\tau_0)/2\{1-\exp(-\tau_0)\}^2$.
%
\end{proposition}

\section{Proof of Theorem \ref{ourlemma1}}


In $\ell_2(\lambda,\beta)$ and $\ell_4(\lambda,\beta)$, $(\lambda_1,\ldots,\lambda_m)^\top $ are not entangled with $\beta$. Therefore, there is no need to develop the minorization functions for them. In the following, we show how to find the minorization functions for $\ell_1(\lambda,\beta)$ and $\ell_3(\lambda,\beta)$.  
Define $u(L_i,X_i)=\sum_{k: t_k\leq L_i}\lambda_{k}+\beta^\top Z_{x_i}(L_i)$, $u(R_i,X_i)=\sum_{k: t_k\leq R_i}\lambda_{k}+\beta^\top Z_{x_i}(R_i)$ and 
$u(L_i, R_i,X_i)=\sum_{k: L_i<t_k\leq R_i}\lambda_{k}+\beta^\top \{Z_{x_i}(R_i)-Z_{x_i}(L_i)\}$. According to 
our model assumption (\ref{eqm1}), 
$u(L_i, X_i)>0$, $u(R_i, X_i)>0$ and $u(L_i, R_i, X_i)>0$ for all $i$.  Now, we can re-write  
 \bse
\ell_1(\lambda,\beta)&=&\sum_{i=1}^n\Delta_{L,i}\log[1-\exp\{-
\sum_{k: t_k\le L_i}\lambda_k-\beta^\top Z_{x_i}(L_i)\}]\\
&=&\sum_{i=1}^n\Delta_{L,i}\log[1-\exp\{-u(L_i, X_i)\}]\\
&=&\sum_{i=1}^n\Delta_{L,i}\left(\log[1-\exp\{-u_0(L_i, X_i)\}]+\log\left[\frac{1-\exp\{-u(L_i, X_i)\}}{1-\exp\{-u_0(L_i, X_i)\}}\right]\right).
\ese
Applying  proposition \ref{prop1} to the second term of the above display  with $\tau=u(L_i,X_i)$ and $\tau_0=u_0(L_i,X_i)$, we obtain
\begin{eqnarray}
\ell_1(\lambda,\beta)&\geq& 
\sum_{i=1}^n\Delta_{L,i}\biggl(
\log[1-\exp\{-u_0(L_i, X_i)\}]+
\{ u(L_i, X_i)-u_0(L_i, X_i)\}
A_1(u_0(L_i, X_i))\nonumber \\
&&-\{ u(L_i, X_i)-u_0(L_i, X_i)\}^2
A_2(u_0(L_i, X_i))+ \log\left\{   \frac{u_0(L_i, X_i)}{u(L_i, X_i)}\right\} +1- \frac{u_0(L_i, X_i)}{u(L_i, X_i)}
\biggl)\nonumber
\\
&=&\sum_{i=1}^n\Delta_{L,i}\Bigg[\{A_1(u_0(L_i,X_i))+2A_2(u_0(L_i,X_i))u_0(L_i,X_i)\}u(L_i,X_i)-A_2(u_0(L_i,X_i))u^2(L_i,X_i)\nonumber \\
&&+\log\left\{\frac{u_0(L_i,X_i)}{u(L_i,X_i)}\right\}-\frac{u_0(L_i,X_i)}{u(L_i,X_i)}
+C_1(u_0(L_i,X_i))\Bigg]\nonumber\\
&=&\sum_{i=1}^n\Delta_{L,i}\Bigg[\{A_1(u_0(L_i,X_i))+2A_2(u_0(L_i,X_i))u_0(L_i,X_i)\}\left(\sum_{k: t_k\leq L_i}\lambda_k+\beta^\top Z_{x_i}(L_i)\right)\nonumber\\
&&-A_2(u_0(L_i,X_i))\left(\sum_{k: t_k\leq L_i}\lambda_k+\beta^\top Z_{x_i}(L_i)\right)^2+\log\left(\frac{u_0(L_i,X_i)}{\sum_{k: t_k\leq L_i}\lambda_k+\beta^\top Z_{x_i}(L_i)}\right)\nonumber\\
&& -\left(\frac{u_0(L_i,X_i)}{\sum_{k: t_k\leq L_i}\lambda_k+\beta^\top Z_{x_i}(L_i)}\right)+C_1(u_0(L_i,X_i))\Bigg],\label{appmyeq1}
\end{eqnarray}
where $C_1(u_0(L_i,X_i))$ is the constant term that only depends on $u_0(L_i,X_i)$, given as
$C_1(u_0(L_i,X_i))=\log[1-\exp\{-u_0(L_i, X_i)\}]-A_1(u_0(L_i,X_i))u_0(L_i,X_i)-A_2(u_0(L_i,X_i))u_0^2(L_i,X_i)+1.$
Next, we look into the following three terms of (\ref{appmyeq1}). First,  \begin{eqnarray*}
-\left(\sum_{t_k\leq L_i}\lambda_k+\beta^\top Z_{x_i}(L_i)\right)^2&=&-\left(\sum_{t_k\le L_i}\frac{\lambda_{k0}}{u_0(L_i,X_i)}\frac{u_0(L_i,X_i)}{\lambda_{k0}}\lambda_k+\frac{\beta_0^\top Z_{x_i}(L_i)}{u_0(L_i,X_i)}\frac{u_0(L_i,X_i)}{\beta_0^\top Z_{x_i}(L_i)}\beta^\top Z_{x_i}(L_i)\right)^2\nonumber\\
&\ge&-\biggl\{\sum_{t_k\leq L_i}\frac{u_0(L_i,X_i)}{\lambda_{k0}}\lambda_k^2+\frac{u_0(L_i,X_i)}{\beta_0^\top Z_{x_i}(L_i)}(\beta^\top Z_{x_i}(L_i))^2\biggl\},
\end{eqnarray*}
where, the inequality is obtained by applying  Jensen's inequality on the concave function $f(x)=-x^2$ and noting that  
$\sum_{k: t_k\le L_i}\lambda_{k0}/u_0(L_i,X_i)+
\beta_0^\top Z_{x_i}(L_i)/u_0(L_i,X_i)=1$. 
Second, applying the standard inequality $\log(x)\ge1-1/x$ for any generic $x>0$, we have 
\bse
\log\left(\frac{u_0(L_i,X_i)}{\sum_{t_k\leq L_i}\lambda_k+\beta^\top Z_{x_i}(L_i)}\right)\geq 1-\frac{\sum_{t_k\leq L_i}\lambda_k+\beta^\top Z_{x_i}(L_i)}{u_0(L_i,X_i)},
\ese
and third,
\bse
-\frac{u_0(L_i,X_i)}{\sum_{t_k\leq L_i}\lambda_k+\beta^\top Z_{x_i}(L_i)}&=&-u_0(L_i,X_i)\biggl\{\sum_{t_k\le L_i}\frac{\lambda_{k0}}{u_0(L_i,X_i)}\frac{u_0(L_i,X_i)}{\lambda_{k0}}\lambda_k\\
&&+\frac{\beta_0^\top Z_{x_i}(L_i)}{u_0(L_i,X_i)}\frac{u_0(L_i,X_i)}{\beta_0^\top Z_{x_i}(L_i)}\beta^\top Z_{x_i}(L_i)\biggl\}^{-1}\\
&\ge &-  \biggl[ \sum_{t_k\le L_i} \frac{\lambda_{k0}^2}{u_0(L_i,X_i)}\lambda_k^{-1}+\frac{\{\beta_0^\top Z_{x_i}(L_i)\}^2}{u_0(L_i,X_i)}\{\beta^\top Z_{x_i}(L_i)\}^{-1}\biggl],
\ese
where, the last inequality is obtained by applying Jensen's inequality on the concave function $f(x)=-1/x$, and noting that  $\sum_{k: t_k\le L_i}\lambda_{k0}/u_0(L_i,X_i)+
\beta_0^\top Z_{x_i}(L_i)/u_0(L_i,X_i)=1$. Then, applying the last three
inequalities in (\ref{appmyeq1}), we obtain $\ell_1(\lambda,\beta)\ge
\ell_{1,\dagger}(\lambda,\beta|\lambda_0,\beta_0)\equiv\sum_{k=1}^m\mathcal{M}_{1,1,k}(\lambda_k|\lambda_0,\beta_0)+\mathcal{M}_{1,2}(\beta|\lambda_0,\beta_0)+\mathcal{M}_{1,3}(\lambda_0,\beta_0)$,
where for $k=1, \dots, m$, 
\bse
\mathcal{M}_{1,1,k}(\lambda_k|\lambda_0,\beta_0)
&=&\sum_{i=1}^n\Delta_{L,i}\Bigg[\{A_1(u_0(L_i,X_i))+2A_2(u_0(L_i,X_i))u_0(L_i,X_i)\}\lambda_k\\
&&-A_2(u_0(L_i,X_i))\left\{\frac{u_0(L_i,X_i)}{\lambda_{k0}}\right\}\lambda_k^2-\frac{\lambda_k}{u_0(L_i,X_i)}-\frac{\lambda_{k0}^2}{u_0(L_i,X_i)}\lambda_k^{-1}\Bigg]I(t_k\leq L_i),
\ese
\bse
\mathcal{M}_{1,2}(\beta|\lambda_0,\beta_0)
&=&\sum_{i=1}^n\Delta_{L,i}\Bigg[\{A_1(u_0(L_i,X_i))+2A_2(u_0(L_i,X_i))u_0(L_i,X_i)\}\beta^\top Z_{x_i}(L_i)\\
&&-A_2(u_0(L_i,X_i))\frac{u_0(L_i,X_i)}{\beta_0^\top Z_{x_i}(L_i)}\{\beta^\top Z_{x_i}(L_i)\}^2-\frac{\beta^\top Z_{x_i}(L_i)}{u_0(L_i,X_i)}\\
&&-\frac{\{\beta_0^\top Z_{x_i}(L_i)\}^2}{u_0(L_i,X_i)}\{\beta^\top Z_{x_i}(L_i)\}^{-1}\Bigg],
\ese
and 
$
\mathcal{M}_{1,3}(\lambda_0,\beta_0)=\sum_{i=1}^n\Delta_{L,i}\{\log[1-\exp\{-u_0(L_i, X_i)\}]-A_1(u_0(L_i,X_i))u_0(L_i,X_i)-A_2(u_0(L_i,X_i))$ $ u_0^2(L_i,X_i)+1\}
$. 
Next, consider finding the minorization function for $\ell_3(\lambda,\beta)$. Here, we use the same techniques as finding the minorization function for $\ell_1(\lambda,\beta)$. Note, 
\bse
\ell_3(\lambda,\beta)
&=&\sum_{i=1}^n\Delta_{I,i}\log\left(1-\exp\left[-\sum_{k: L_i< t_k\leq R_i}\lambda_k-\beta^\top \{Z_{x_i}(R_i)-Z_{x_i}(L_i)\}\right]\right)\\
&=&\sum_{i=1}^n\Delta_{I,i}\log[1-\exp\{-u(L_i,R_i, X_i)\}]\\
&=&\sum_{i=1}^n\Delta_{I,i}\left(\log[1-\exp\{-u_0(L_i,R_i, X_i)\}]+\log\left[\frac{1-\exp\{-u(L_i,R_i, X_i)\}}{1-\exp\{-u_0(L_i,R_i, X_i)\}}\right]\right).
\ese
Now applying proposition \ref{prop1} to the second term of the above display with $\tau=u(L_i,R_i,X_i)$ and $\tau_0=u_0(L_i,R_i,X_i)$, we obtain
\begin{eqnarray}
\ell_3(\lambda,\beta)&\geq& 
\sum_{i=1}^n\Delta_{I,i}\biggl(
\log[1-\exp\{-u_0(L_i,R_i X_i)\}]+
\{ u(L_i,R_i, X_i)-u_0(L_i,R_i, X_i)\}
A_1(u_0(L_i,R_i, X_i))\nonumber \\
&&-\{ u(L_i,R_i, X_i)-u_0(L_i,R_i, X_i)\}^2
A_2(u_0(L_i,R_i, X_i))\nonumber\\
&&+ \log\left\{   \frac{u_0(L_i,R_i, X_i)}{u(L_i,R_i, X_i)}\right\} +1- \frac{u_0(L_i,R_i, X_i)}{u(L_i,R_i, X_i)}
\biggl)\nonumber
\\
&=&\sum_{i=1}^n\Delta_{I,i}\Bigg[\{A_1(u_0(L_i,R_i,X_i))+2A_2(u_0(L_i,R_i,X_i))u_0(L_i,R_i,X_i)\}u(L_i,R_i,X_i)\nonumber\\
&&-A_2(u_0(L_i,R_i,X_i))u^2(L_i,R_i,X_i)\nonumber \\
&&+\log\left\{\frac{u_0(L_i,R_i,X_i)}{u(L_i,R_i,X_i)}\right\}-\frac{u_0(L_i,R_i,X_i)}{u(L_i,R_i,X_i)}
+C_1(u_0(L_i,R_i,X_i))\Bigg]\nonumber\\
&=&\sum_{i=1}^n\Delta_{I,i}\Bigg[\{A_1(u_0(L_i,R_i,X_i))\nonumber\\
&&+2A_2(u_0(L_i,R_i,X_i))u_0(L_i,R_i,X_i)\}\left(\sum_{k: L_i<t_k\leq R_i}\lambda_k+\beta^\top (Z_{x_i}(R_i)-Z_{x_i}(L_i))\right)\nonumber\\
&&-A_2(u_0(L_i,R_i,X_i))\left(\sum_{k: L_i<t_k\leq R_i}\lambda_k+\beta^\top (Z_{x_i}(R_i)-Z_{x_i}(L_i))\right)^2\nonumber\\
&&+\log\left(\frac{u_0(L_i,R_i,X_i)}{
\sum_{k: L_i<t_k\leq R_i}\lambda_k+\beta^\top (Z_{x_i}(R_i)-Z_{x_i}(L_i))}
\right)\nonumber\\
&& -\left(\frac{u_0(L_i,R_i,X_i)}{\sum_{k: L_i<t_k\leq R_i}\lambda_k+\beta^\top (Z_{x_i}(R_i)-Z_{x_i}(L_i))}\right)+C_1(u_0(L_i,R_i,X_i))\Bigg]\label{appmyeq1_l3}
\end{eqnarray}
where, $C_1(u_0(L_i,R_i,X_i))$ is the constant term that only depends on $u_0(L_i,R_i,X_i)$, given by 
$
C_1(u_0(L_i,R_i,X_i))=\log[1-\exp\{-u_0(L_i,R_i, X_i)\}]-A_1(u_0(L_i,R_i,X_i))u_0(L_i,R_i,X_i)-A_2(u_0(L_i,R_i,X_i))u_0^2(L_i,R_i,X_i)+1.
$
Similarly, we have the following three inequalities,
\begin{eqnarray*}
&-&\left(\sum_{L_i<t_k\leq R_i}\lambda_k+\beta^\top (Z_{x_i}(R_i)-Z_{x_i}(L_i))\right)^2\\
&=&-\left(\sum_{L_i<t_k\le R_i}\frac{\lambda_{k0}}{u_0(L_i,R_i,X_i)}\frac{u_0(L_i,R_i,X_i)}{\lambda_{k0}}\lambda_k\right.\\
&&\left.+\frac{\beta_0^\top (Z_{x_i}(R_i)-Z_{x_i}(L_i))}{u_0(L_i,R_i,X_i)}\frac{u_0(L_i,R_i,X_i)}{\beta_0^\top (Z_{x_i}(R_i)-Z_{x_i}(L_i))}\beta_0^\top (Z_{x_i}(R_i)-Z_{x_i}(L_i))\right)^2\nonumber\\
&\ge&-\biggl\{\sum_{L_i<t_k\leq R_i}\frac{u_0(L_i,R_i,X_i)}{\lambda_{k0}}\lambda_k^2+\frac{u_0(L_i,R_i,X_i)}{\beta_0^\top (Z_{x_i}(R_i)-Z_{x_i}(L_i))}(\beta^\top (Z_{x_i}(R_i)-Z_{x_i}(L_i)))^2\biggl\},
\end{eqnarray*}
\bse
\log\left(\frac{u_0(L_i,R_i,X_i)}{\sum_{L_i<t_k\leq R_i}\lambda_k+\beta^\top (Z_{x_i}(R_i)-Z_{x_i}(L_i))}\right)\geq 1-\frac{\sum_{L_i<t_k\leq R_i}\lambda_k+\beta^\top (Z_{x_i}(R_i)-Z_{x_i}(L_i))}{u_0(L_i,R_i,X_i)},
\ese
and
\bse
&-&\frac{u_0(L_i,R_i,X_i)}{\sum_{L_i<t_k\leq R_i}\lambda_k+\beta^\top (Z_{x_i}(R_i)-Z_{x_i}(L_i))}\\
&=&-u_0(L_i,R_i,X_i)\biggl\{\sum_{L_i<t_k\leq R_i}\frac{\lambda_{k0}}{u_0(L_i,R_i,X_i)}\frac{u_0(L_i,R_i,X_i)}{\lambda_{k0}}\lambda_k\\
&&+\frac{\beta_0^\top (Z_{x_i}(R_i)-Z_{x_i}(L_i))}{u_0(L_i,R_i,X_i)}\frac{u_0(L_i,R_i,X_i)}{\beta_0^\top (Z_{x_i}(R_i)-Z_{x_i}(L_i))}\beta^\top (Z_{x_i}(R_i)-Z_{x_i}(L_i))\biggl\}^{-1}\\
&\ge &-  \biggl[ \sum_{L_i<t_k\le R_i} \frac{\lambda_{k0}^2}{u_0(L_i,R_i,X_i)}\lambda_k^{-1}+\frac{\{\beta_0^\top (Z_{x_i}(R_i)-Z_{x_i}(L_i))\}^2}{u_0(L_i,R_i,X_i)}\{\beta^\top (Z_{x_i}(R_i)-Z_{x_i}(L_i))\}^{-1}\biggl],
\ese
where, the first and the third inequalities are obtained by applying Jensen's inequality on the  concave function $f(x)=-x^2$ and $f(x)=-1/x$, respectively, and the second inequality is obtained by applying the standard inequality $\log(x)\ge1-1/x$.
Applying the above two inequalities in (\ref{appmyeq1_l3}),  we obtain  $\ell_3(\lambda,\beta)\ge\ell_{3,\dagger}(\lambda,\beta|\lambda_0,\beta_0)\equiv\sum_{k=1}^m\mathcal{M}_{3,1,k}(\lambda_k|\lambda_0,\beta_0)+\mathcal{M}_{3,2}(\beta|\lambda_0,\beta_0)+\mathcal{M}_{3,3}(\lambda_0,\beta_0)$,
where
\bse
\mathcal{M}_{3,1,k}(\lambda_k|\lambda_0,\beta_0)
&=&\sum_{i=1}^n\Delta_{I,i}\Bigg[\{A_1(u_0(L_i,R_i,X_i))+2A_2(u_0(L_i,R_i,X_i))u_0(L_i,R_i,X_i)\}\lambda_k\\
&&-A_2(u_0(L_i,R_i,X_i))\left\{\frac{u_0(L_i,R_i,X_i)}{\lambda_{k0}}\right\}\lambda_k^2\\
&&-\frac{\lambda_k}{u_0(L_i,R_i,X_i)}-\frac{\lambda_{k0}^2}{u_0(L_i,R_i,X_i)}\lambda_k^{-1}\Bigg]I(L_i< t_k\le R_i),\quad k=1,\ldots,m,
\ese
\bse
\mathcal{M}_{3,2}(\beta|\lambda_0,\beta_0)
&=&\sum_{i=1}^n\Delta_{I,i}\Bigg(\{A_1(u_0(L_i,R_i,X_i))+2A_2(u_0(L_i,R_i,X_i))u_0(L_i,R_i,X_i)\}\\
&&\times \beta^\top \{Z_{x_i}(R_i)-Z_{x_i}(L_i)\}
-A_2(u_0(L_i,R_i,X_i))\frac{u_0(L_i,R_i,X_i)[\beta^\top \{Z_{x_i}(R_i)-Z_{x_i}(L_i)\}]^2}{\beta_0^\top \{Z_{x_i}(R_i)-Z_{x_i}(L_i)\}}\\
&&-\frac{\beta^\top \{Z_{x_i}(R_i)-Z_{x_i}(L_i)\}}{u_0(L_i,R_i,X_i)}
-\frac{[\beta_0^\top \{Z_{x_i}(R_i)-Z_{x_i}(L_i)\}]^2}{u_0(L_i,R_i,X_i) \beta^\top \{Z_{x_i}(R_i)-Z_{x_i}(L_i)\}   }
\Bigg),
\ese
and
\bse
\mathcal{M}_{3,3}(\lambda_0,\beta_0)&=&\sum_{i=1}^n\Delta_{I,i}\Bigg[\log\left\{1-\exp\left(-\left[\sum_{L_i< t_k\leq R_i}\lambda_k+\beta^\top \{Z_{x_i}(R_i)-Z_{x_i}(L_i)\}\right]\right)\right\}\\
&&-A_1(u_0(L_i,R_i,X_i))u_0(L_i,R_i,X_i)-A_2(u_0(L_i,R_i,X_i))u_0^2(L_i,R_i,X_i)+1\Bigg].
\ese
%
%
Finally, we obtain 
 \bse 
 \ell(\lambda, \beta)&=&\ell_1(\lambda, \beta)+
 \ell_2(\lambda, \beta)+\ell_3(\lambda, \beta)\\
 &\geq&  
 \ell_{\dagger}(\lambda,\beta|\lambda_0,\beta_0)\\
 &
 \equiv&  \ell_{1,\dagger}(\lambda,\beta|\lambda_0,\beta_0)+\ell_2(\lambda, \beta)+\ell_{3,\dagger}(\lambda,\beta|\lambda_0,\beta_0)\\
 &=& \sum_{k=1}^m\mathcal{M}_{1,1,k}(\lambda_k|\lambda_0,\beta_0)+\mathcal{M}_{1,2}(\beta|\lambda_0,\beta_0)+\mathcal{M}_{1,3}(\lambda_0,\beta_0)+\ell_2(\lambda, \beta)\\
 &&+\sum_{k=1}^m\mathcal{M}_{3,1,k}(\lambda_k|\lambda_0,\beta_0)+\mathcal{M}_{3,2}(\beta|\lambda_0,\beta_0)+\mathcal{M}_{3,3}(\lambda_0,\beta_0)\\
 &\equiv &
 \sum_{k=1}^m\mathcal{M}_{1,k}(\lambda_k|\lambda_0,\beta_0)+\mathcal{M}_2(\beta|\lambda_0,\beta_0)+\mathcal{M}_3(\lambda_0,\beta_0),
 \ese 
where 
$ 
\mathcal{M}_{1,k}(\lambda_k|\lambda_0,\beta_0)
= \mathcal{M}_{1,1,k}(\lambda_k|\lambda_0,\beta_0)+ \mathcal{M}_{3,1,k}(\lambda_k|\lambda_0,\beta_0)- \lambda_k\sum^n_{i=1} \Delta_{I, i}I(t_k\leq L_i)
$, 
$  
\mathcal{M}_{2}(\beta|\lambda_0,\beta_0)
= \mathcal{M}_{1, 2}(\beta|\lambda_0,\beta_0)+ \mathcal{M}_{3, 2}(\beta|\lambda_0,\beta_0)-
\sum_{i=1}^n\Delta_{I,i}\beta^\top Z_{x_i}(L_i)$, 
and $\mathcal{M}_3(\lambda_0,\beta_0)
= \mathcal{M}_{1, 3}(\lambda_0,\beta_0)
+\mathcal{M}_{3, 3}(\lambda_0,\beta_0)$.

\newpage

\end{document}